\documentclass[journal,article,accept,moreauthors,pdftex,12pt,a4paper]{mdpi} 

\usepackage{epsf,epsfig,subfigure,graphicx,amsmath,amssymb}

\def\to{\rightarrow}

\def\bi{\begin{itemize}}
\def\ei{\end{itemize}}

\def\ta{\tilde a}

\def\sps1ap{SPS1a$^\prime$}
\def\c1p{C1$^\prime$}

\def\tb{\tilde b}

\def\tst{\tilde t}

\def\tg{\tilde g}

\def\tw{\widetilde W}
\def\tz{\widetilde Z}
\def\be{\begin{equation}}  
\def\ee{\end{equation}}  
\def\bea{\begin{eqnarray}}  
\def\eea{\end{eqnarray}}  
\def\beas{\begin{eqnarray*}}  
\def\eeas{\end{eqnarray*}}

\lastpage{x}
\doinum{10.3390/------}
\pubvolume{xx}
\pubyear{2015}
\history{Received: xx / Accepted: xx / Published: xx}

\usepackage{graphicx, epsfig, color}

\Title{Supersymmetry with radiatively-driven naturalness:\\                     
implications for WIMP and axion searches}

\Author{Kyu Jung Bae$^{1}$, Howard Baer$^{1,}$*, Vernon Barger$^{2}$,
Michael R. Savoy$^{1}$ and Hasan Serce$^{1}$
}

\address{%
$^{1}$ Dept. of Physics and Astronomy, University of Oklahoma, Norman, OK 73019, USA\\
$^{2}$ Dept. of Physics, University of Wisconsin, Madison, WI 53706, USA}

\corres{baer@nhn.ou.edu, 405-325-3961 ext 36315}

\abstract{By insisting on naturalness in both the electroweak and QCD sectors of the
MSSM, the portrait for dark matter production is seriously modified 
from the usual WIMP miracle picture. In SUSY models with radiatively-driven naturalness 
(radiative natural SUSY or RNS)
which include a DFSZ-like solution to the strong CP and SUSY $\mu$ problems, 
dark matter is expected to be an admixture of both axions and higgsino-like WIMPs. 
The WIMP/axion abundance calculation requires simultaneous solution of a 
set of coupled Boltzmann equations which describe quasi-stable axinos and saxions.
In most of parameter space, axions make up the dominant contribution of dark matter
although regions of WIMP dominance also occur.
We show the allowed range of PQ scale $f_a$ and compare to the values expected
to be probed by the ADMX axion detector in the near future. 
We also show WIMP detection rates which are
suppressed from usual expectations because now WIMPs comprise only a fraction of the 
total dark matter. 
Nonetheless, ton-scale noble liquid detectors should be able to probe the entirety
of RNS parameter space. Indirect WIMP detection rates are less propitious since they
are reduced by the square of the depleted WIMP abundance.
}


\begin{document}

\newpage
\section{Introduction}

\begin{figure}[tbp]
\begin{center}
\includegraphics[height=0.25\textheight]{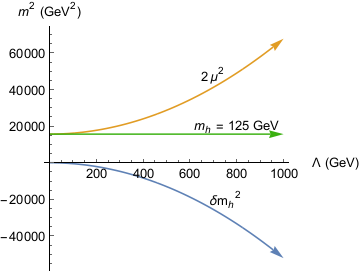}
\caption{Plot of measured Higgs mass squared along with radiative 
correction and tree-level term $2\mu^2$. The latter term 
is adjusted (fine-tuned) to guarantee that $m_h=125$ GeV.
\label{fig:mh2}}
\end{center}
\end{figure}

The discovery of the Higgs boson~\cite{atlas_h,cms_h} with mass $m_h=125.15\pm 0.24$ GeV 
was a great triumph but it brings with it a conundrum: 
how is it that scalar fields can actually occur in nature?
The problem lies in the radiative corrections to their masses: they are quadratically divergent 
in the energy circulating in the loop diagrams. 
Since quantum mechanics requires one to sum over a complete 
set of states, those states with the highest energies bring large quantum corrections which must be
compensated by adjusting bare mass terms to maintain the measured value of $m_h$. 
The situation is depicted in Fig.~\ref{fig:mh2}: here we take the SM Higgs potential 
as $V=-\mu_h^2|h|^2+\lambda_h |h|^4$ where $m_h^2=2\mu_h^2+\delta m_h^2$ 
and $m_h^2({\rm tree})=2\mu_h^2$.
Requiring the quantum corrections not exceed the bare mass 
(similar to the Gaillard-Lee~\cite{GL} requirement on $\Delta m_K^2$ which predicted the charm quark mass) 
implies the Standard Model to only be valid at energy scales $Q\lesssim \Lambda\sim 1$ TeV.
These quadratic divergences-- which are endemic to scalar quantum fields-- led some physicists 
to ponder whether fundamental scalar fields could really occur in nature~\cite{Susskind:1978ms}.

The solution to the above SM naturalness problem was very conservative: expand the fundamental 
$4-D$ spacetime symmetry structure which underlies quantum field theory to its most general 
structure including graded Lie-algebras~\cite{susyreviews,wss}. 
The expanded symmetry group-- called supersymmetry or SUSY for short--
provided once and for all the necessary structure so that scalar field quadratic 
divergences completely cancelled. 
Akin to the doubling of particle spectra which occurred when Dirac included 
Lorentz symmetry into quantum mechanics, SUSY also requires an approximate doubling: 
under SUSY, for every boson there is a fermion state and vice versa. 
Since we see {\it e.g.} no bosonic electrons with the same mass as electron 
(similar arguments apply to other SM particle states), SUSY must be a broken symmetry.
To stabilize the weak scale, it is expected that SUSY breaking is characterized 
by soft SUSY breaking terms of weak scale magnitude. 
In fact, in models bases on local supersymmetry (supergravity or SUGRA), 
the breakdown of SUSY must occur in a ``hidden sector'' of the model to maintain
phenomenological viability~\cite{sugra}. 
Taking the limit of $M_P\to\infty$ while keeping the gravitino mass $m_{3/2}$ fixed, 
one calculates the soft terms~\cite{soft} as multiples of $m_{3/2}$ where $m_{3/2}\sim m_{\rm hidden}^2/M_P$. 
Here $M_P=2.4\times10^{18}$ GeV is the reduced Planck mass.
A hidden sector mass scale $m_{\rm hidden}\sim 10^{10}$ GeV gives
rise to a weak scale of $\sim 100$ GeV. 

From the above arguments, we arrive at the rough expectation that the new matter particles
should inhabit the energy scale $Q\sim 100-1000$ GeV. 
Lest one think the above construct is the product of an overly active imagination of theorists, 
we remark that SUSY is supported by three disparate sets of measurements:
\begin{itemize}
\item The measured values of the three gauge couplings, when extrapolated to $m_{\rm GUT}\simeq 2\times 10^{16}$ GeV, 
very nearly meet at a point~\cite{gauge}, as expected in simple unified theories.
\item The measured value of the top quark, $m_t=173.2$ GeV, is in just the right range 
to drive the up-Higgs soft mass $m_{H_u}^2$ to negative values, 
causing the required breakdown of electroweak symmetry~\cite{rewsb}.
\item The measured value of the newly discovered higgs boson, $m_h\simeq 125$ GeV, 
falls squarely within the narrow window $m_h\sim 115-135$ GeV 
of SUSY requirements which was expected from the pre-LHC era~\cite{mhiggs}. 
In contrast, in the SM the Higgs mass could lie anywhere in the $115-800$ GeV mass range.
\end{itemize}
In addition, SUSY-- as embodied by the MSSM-- carries with it several dark matter 
candidates~\cite{dmreviews} and several baryogenesis mechanisms~\cite{baryoreviews} 
whereas the SM contains neither.

In spite of these successes, many authors have proclaimed weak-scale SUSY to be in a state 
of crisis~\cite{crisis}. 
While SUSY solves the big hierarchy problem involving quadratic divergences~\cite{witten}, 
there is a growing Little Hierarchy problem~\cite{LH} typified by the increasing gap 
between the $W$, $Z$ and $h$ masses clustered all around $\sim 100$ GeV, 
and the apparent mass scale of SUSY particles which are seemingly in the multi-TeV range. 
Presently, LHC8 with 20 fb$^{-1}$ of data requires $m_{\tg}\gtrsim 1.3$ TeV in the case of 
heavy squark masses and $m_{\tg}\gtrsim 1.8$ TeV in the case of comparable squark masses. 
Furthermore, the value of $m_h\sim 125$ GeV requires radiative corrections from 
top-squarks in the tens of TeV range for small top-squark mixing
(although few-TeV top squarks are allowed for large mixing induced by 
trilinear $A$ terms~\cite{h125}).
The lore is that as the mass scale for the soft terms increases, 
then one must increasingly fine-tune parameters to maintain $m_{W,Z,h}\sim 100$ GeV.
Since large fine-tuning usually indicates some {\it pathology} in any theoretical construct, 
a number of authors have questioned whether SUSY as we know it is gradually becoming excluded~\cite{crisis}: 
if so, then new ideas for physics beyond the Standard model are required.

In the following Section~\ref{sec:nat}, we shall refute this point of view. 
While we shall conclude that many SUSY models
are indeed fine-tuned-- including the paradigm mSUGRA/CMSSM model--
we will find that models characterized by radiatively-driven naturalness~\cite{ltr,rns} 
(radiatively-driven natural SUSY or RNS) are allowed with modest fine-tunings only at 
the 10\% level.
Radiatively-driven naturalness occurs in SUSY models with non-universality of 
Higgs soft terms (as in the NUHM2 model~\cite{nuhm2}).
RNS models are characterized by the presence of light higgsinos with mass $\mu\sim 100-200$ GeV, 
the closer to $m_Z$ the better. 
The lightest SUSY particle is a candidate for dark matter and is then a higgsino-like WIMP.

We proceed to examine the consequences of RNS for dark matter. 
In Sec.~\ref{sec:qcd}, we require that naturalness occurs also 
in the QCD sector of the MSSM. This brings to bear the QCD axion albeit as one 
element of a axion supermultiplet
containing also a spin-1/2 $R$-parity odd axino $\ta$ and a spin-0 $R$-parity even 
saxion field $s$.
The dark matter then consists of a combination of both axions and higgsino-like WIMPs.
In Sec.~\ref{sec:axions}, we present calculations of the expected abundance of axions and WIMPs 
in RNS SUSY. 
We display the range in PQ breaking scale $f_a$ which is accessible to axion search 
experiments like ADMX~\cite{admx}.
In Sec.~\ref{sec:wimps}, we examine updated prospects for WIMP detection in RNS.
While higgsinos may comprise as little as 5-10\% of the total dark matter abundance, they
should nonetheless be detectable by ton-scale WIMP direct detection experiments 
owing to their large couplings to the Higgs boson $h$. 
Indirect WIMP detection seems less propitious since the
detection rate is proportional to the square of the reduced WIMP abundance.
We conclude in Sec.~\ref{sec:conclude}.


\section{Measuring naturalness in SUSY theories}
\label{sec:nat}

Any serious discussion of naturalness requires the definition of some measure. 
But first, an important point to be made is that {\it any} quantity can look 
fine-tuned if one splits it into {\it dependent} pieces.
By re-writing an observable ${\cal O}$ as ${\cal O}+b -b$ 
and allowing $b$ to be large, the quantity might look fine-tuned. 
In this trivial example, however, combining dependent contributions into independent units 
($b-b=0$) obviously erases the presumed source of fine-tuning.
To avoid such pitfalls, a simple fine-tuning rule has been proposed~\cite{dew}:
\begin{quotation}
\noindent When evaluating fine-tuning, it is not permissible to claim fine-tuning 
of {\it dependent} quantities one against another.
\end{quotation}

\subsection{Simple electroweak fine-tuning}

The simplest relation between the weak scale and the soft SUSY breaking parameters comes
from minimizing the scalar potential of the MSSM to determine the vacuum expectation values (VEVs)~\cite{wss}.
The first minimization condition allows one to trade the bilinear soft term $B$ for 
the more convenient ratio of VEVs $\tan\beta \equiv v_u/v_d$. 
The second condition is given by
\bea
\frac{m_Z^2}{2} &=& \frac{(m_{H_d}^2+\Sigma_d^d)-(m_{H_u}^2+\Sigma_u^u)\tan^2\beta}{(\tan^2\beta -1)}
-\mu^2\\
&\simeq &-m_{H_u}^2-\mu^2-\Sigma_u^u
\label{eq:mzs}
\eea
where $m_{H_u}^2$ and $m_{H_d}^2$ are the {\it weak scale} soft SUSY breaking Higgs masses, $\mu$
is the {\it supersymmetric} higgsino mass term and $\Sigma_u^u$ and $\Sigma_d^d$ contain
an assortment of loop corrections to the effective potential (for a listing, see Ref.~\cite{rns}).
For naturalness, we require no large unnatural cancellations between independent terms on the
right-hand-side of Eq.~\ref{eq:mzs}.
For instance, if $m_{H_u}^2$ is driven to multi-TeV negative values at the weak scale, 
then the {\it completely unrelated} value of $\mu^2$ is required to be multi-TeV positive with such
high precision as to yield a $Z$ mass of just $91.2$ GeV. 
This fine-tuning occurs on a daily basis by users of SUSY spectrum generator 
tools~\cite{isajet,suspect,softsusy,spheno}, but it is hidden in the computer code.
While such tuning is logically possible, 
the overall scenario seems highly implausible, or highly unnatural 
(in this case, the $Z$ mass would naturally be expected occur in the multi-TeV range).

The quantity $\Delta_{\rm EW}$ measures this implausibility by 
comparing the largest contribution on the right-hand-side of Eq.~\ref{eq:mzs} 
to the value of $m_Z^2/2$. If they are comparable, then no
unnatural fine-tunings are required to generate $m_Z=91.2$ GeV.

The main requirements for EW naturalness can then be read off from Eq.~\ref{eq:mzs}.
They are the following: 
\begin{itemize}
\item $|\mu |\sim 100-200$ GeV (the closer to $m_Z$ the 
better)~\cite{Chan:1997bi,Barbieri:2009ev,hgsno}. 
We note here that the lower bound on $\mu \gtrsim 100$ GeV comes from accommodating LEP2 limits 
from chargino pair production searches. A low value of $\Delta_{\rm EW}$ yields an upper bound on 
$|\mu |$ depending on how much fine-tuning one is willing to tolerate. 
A value $\Delta_{\rm EW}<10$ (or $\Delta_{\rm EW}^{-1}> 10\%$) for fine-tuning implies $|\mu |<200$ GeV.
\item The value of $m_{H_u}^2$ is driven radiatively to small, 
and not large, negative values~\cite{ltr,rns}.
In the mSUGRA/CMSSM model, this occurs in the hyperbolic branch/focus point (HB/FP) 
region~\cite{hb_fp}.
However, the rather large value of $m_h$ requires a large trilinear $A_0$ parameter. 
Such a large trilinear pushes the HB/FP out to typically $m_0\sim 10-30$ TeV~\cite{sugpaper}. 
At such high $m_0$, then the top squark contributions
$\Sigma_u^u(\tst_{1,2})$ become large and again one is fine-tuned.
Alternatively, in models where the Higgs soft terms are {\it non-universal}, 
such as in the two-extra parameter non-universal Higgs model NUHM2~\cite{nuhm2}, 
it is possible to have small $\mu$ for any $m_0$ value  by simply raising 
the GUT scale value of $m_{H_u}({\rm GUT})\sim (1.3-2)m_0$.
\item The top squark contributions to the radiative corrections $\Sigma_u^u(\tst_{1,2})$ 
can become large for stops in the multi-TeV region. 
However, the radiative corrections are minimized for {\it highly mixed} (large $A_0$) 
top squarks~\cite{ltr}. 
This latter condition  also lifts the Higgs mass to $m_h\sim 125$ GeV.
\end{itemize}
The measure $\Delta_{\rm EW}$ is pre-programmed in the Isajet SUSY spectrum generator 
called Isasugra~\cite{isajet,isasugra}.

One advantage of $\Delta_{\rm EW}$ is that-- within the context of the MSSM-- it is (as discussed in Ref.~\cite{rns}) 
1. {\it model-independent}: if a weak scale spectrum is generated
within the pMSSM or via some high scale constrained model, one obtains exactly the
same value of naturalness. 
Other virtues of $\Delta_{\rm EW}$ are
that it is: 2. the most conservative of the proposed measures,
3. in principle measureable, 4. unambiguous, 5. predictive, 6. falsifiable 
and 7. simple to calculate.

The principle criticism of $\Delta_{\rm EW}$ is that-- since it involves only weak scale parameters-- 
it may not display the sensitivity of the weak scale to variations in high scale parameters. 
Below we discuss two competing measures, $\Delta_{\rm HS}$ and $\Delta_{\rm BG}$. 
Typically, these latter two measures are implemented in violation of the fine-tuning rule. 
If implemented in accord with the fine-tuning rule, then both essentially reduce  to $\Delta_{\rm EW}$.
In this case, $\Delta_{\rm EW}$ portrays the entirety of electroweak naturalness even 
including high scale physics.

\subsubsection{Large-log measure  $\Delta_{\rm HS}$}

The Higgs mass fine-tuning measure, $\Delta_{\rm HS}$, compares the radiative correction of the 
$m_{H_u}^2$ soft term, $\delta m_{H_u}^2$,  to the physical Higgs mass $m_h^2$:
\bea
\Delta_{\rm HS}&=& \delta m_{H_u}^2/(m_h^2/2)\ \ \ {\rm where}\\
m_h^2&\sim &\mu^2+m_{H_u}^2(\Lambda )+\delta m_{H_u}^2 .
\label{eq:mhs}
\eea
If we assume the MSSM is valid up to some high energy scale $\Lambda$ (which may be as high as $m_{\rm GUT}$ or even $M_P$), 
then the value of $\delta m_{H_u}^2$ can be found by integrating the renormalization group equation (RGE):
\be
\frac{dm_{H_u}^2}{dt}=\frac{1}{8\pi^2}\left(-\frac{3}{5}g_1^2M_1^2-3g_2^2M_2^2+\frac{3}{10}g_1^2 S+3f_t^2 X_t\right)
\label{eq:mHu}
\ee
where $t=\ln (Q^2/Q_0^2)$,
$S=m_{H_u}^2-m_{H_d}^2+Tr\left[{\bf m}_Q^2-{\bf m}_L^2-2{\bf m}_U^2+{\bf m}_D^2+{\bf m}_E^2\right]$
and $X_t=m_{Q_3}^2+m_{U_3}^2+m_{H_u}^2+A_t^2$.
By neglecting gauge terms and $S$ ($S=0$ in models with scalar soft term universality 
but can be large in models with non-universality),
and also neglecting the $m_{H_u}^2$ contribution to $X_t$ and the fact that $f_t$ and the soft terms
evolve under $Q^2$ variation, 
a simple expression may be obtained by integrating from $m_{\rm SUSY}$ to the cutoff $\Lambda$:
\be
\delta m_{H_u}^2 \sim -\frac{3f_t^2}{8\pi^2}(m_{Q_3}^2+m_{U_3}^2+A_t^2)\ln\left(\Lambda^2/m_{\rm SUSY}^2 \right) .
\label{eq:DBoE}
\ee
Here, we take as usual $m_{\rm SUSY}^2 \simeq m_{\tst_1}m_{\tst_2}$.
By requiring~\cite{kitnom,papucci,brust,Evans:2013jna} 
\be
\Delta_{\rm HS}\lesssim 10
\ee 
then one expects the three third generation squark masses $m_{\tst_{1,2},\tb_1}\lesssim 600$ GeV.
Using the $\Delta_{\rm HS}$ measure of fine-tuning along with $m_h\simeq 125$ GeV, one finds 
some popular SUSY models to be electroweak fine-tuned to 0.1\%~\cite{comp}.

Two problems occur within this approach.
\begin{enumerate}
\item $m_{H_u}^2(\Lambda )$ and $\delta m_{H_u}^2$ are {\it not} independent:
the value of $m_{H_u}^2$ feeds directly into evaluation of $\delta m_{H_u}^2$ via the $X_t$ term:
the larger the value of $m_{H_u}^2(\Lambda )$, then the larger is the cancelling correction $\delta m_{H_u}^2$~\cite{arno}.
It also feeds indirectly into $\delta m_{H_u}^2$ by contributing to the evolution of the
$m_{Q_3}^2$ and $m_{U_3}^2$ terms. 
Thus, the $\Delta_{\rm HS}$ measure as constructed {\it fails} the fine-tuning rule~\cite{dew}.
\item In the SM, the SU(2)$_L\times$U(1)$_Y$ gauge symmetry can be broken at tree
level. However, in the case of SUGRA gauge theories, where SUSY is broken in a hidden sector via the superHiggs mechanism, 
$m_{H_u}^2\sim m_{3/2}^2>0$. Thus, for SUGRA models, electroweak symmetry is not even broken until
one includes radiative corrections. For SUSY models valid up to some high scale $\Lambda\gg m_{\rm weak}$,
the large log in Eq.~\ref{eq:DBoE} is exactly what is required to break EW symmetry in the first place,
radiatively driving $m_{H_u}^2$ to negative values~\cite{rewsb}.
\end{enumerate}
A simple fix for $\Delta_{\rm HS}$ is to {\it combine the dependent terms} into a single quantity. 
Under such a regrouping~\cite{ltr,rns}, 
\be
m_h^2\simeq \mu^2+\left(m_{H_u}^2(\Lambda )+\delta m_{H_u}^2 \right)
\label{eq:mh}
\ee
where now $\mu^2$ and $\left(m_{H_u}^2(\Lambda )+\delta m_{H_u}^2 \right)$ are each independent so each 
should be comparable to $m_h^2$ in order to avoid fine-tuning. 
The large log is still present in $(m_{H_u}^2(\Lambda )+\delta m_{H_u}^2)$, but now cancellations can occur
between the boundary condition and the radiative correction.
The regrouping of contributions to $m_h^2$ leads back to the $\Delta_{\rm EW}$ measure since 
now $(m_{H_u}^2(\Lambda )+\delta m_{H_u}^2) = m_{H_u}^2({\rm weak})$. 

\subsection{The EENZ/BG measure}

The traditional measure, $\Delta_{\rm BG}$, was proposed by Ellis, Enquist, Nanopoulos and Zwirner~\cite{Ellis:1986yg} 
and later investigated more thoroughly by Barbieri and Giudice~\cite{bg}. 
The proposal is that the variation in $m_Z^2$ with respect to high scale parameter variation 
be small:
\be
\Delta_{\rm BG}\equiv \max\left[ c_i\right]\ \ {\rm where}\ \ 
c_i=\left|\frac{\partial\ln m_Z^2}{\partial\ln p_i}\right|
=\left|\frac{p_i}{m_Z^2}\frac{\partial m_Z^2}{\partial p_i}\right|
\label{eq:DBG}
\ee
where the $p_i$ constitute the fundamental parameters of the model.
Thus, $\Delta_{\rm BG}$ measures the fractional change in $m_Z^2$ due to fractional variation in
high scale parameters $p_i$.
The $c_i$ are known as {\it sensitivity coefficients}~\cite{bg}. 

To evaluate $\Delta_{\rm BG}$, we first express $m_Z^2$ in terms of weak scale SUSY parameters 
as in Eq.~\ref{eq:mzs}:
\be
m_Z^2 \simeq -2m_{H_u}^2-2\mu^2 ,
\label{eq:mZsapprox}
\ee
where the partial equality obtains for moderate-to-large $\tan\beta$ values and where we assume for
now the radiative corrections are small.
Next, one needs to know the explicit dependence of 
the weak scale values of $m_{H_u}^2$ and $\mu^2$ on the more fundamental high scale parameters. 
These can be obtained from semi-analytic solutions to the renormalization group equations
for $m_{H_u}^2$ and $\mu^2$ and can be found in Ref.~\cite{munoz}.

The place where the application of $\Delta_{\rm BG}$ can go wrong is in the identification of the
fundamental parameter set $p_i$. 
Usually, the set $p_i$ is taken to be the various soft terms of particular 
effective theories such as  the MSSM, mSUGRA, NUHM2, {\it etc.} which arise from 
integrating out the hidden sector of the underlying SUGRA theory. 
In these effective theories, variation of the soft SUSY breaking parameters 
allows for a wide range of possibilities for the (unknown) hidden sector 
and the dynamics of SUSY breaking.
However, recall that in SUGRA gauge theories with SUSY broken in a hidden sector, 
all soft parameters are actually computed as multiples of the gravitino mass $m_{3/2}$. 
This means that for any given hidden sector, 
the soft terms are all {\it correlated}: 
if one increases the value of $m_{3/2}$, then all soft terms increase in magnitude accordingly: 
{\it i.e.} in SUGRA they are {\it not} independent. 
Combining the contributions of the {\it dependent} high-scale soft terms to $m_Z^2$, we arrive 
at the simple high scale relation
\bea
m_Z^2&\sim &-2\mu^2({\rm weak})-2m_{H_u}^2({\rm weak}) \nonumber\\
&\sim & -2\mu^2({\rm GUT})+a\cdot m_{3/2}^2 .
\label{eq:mzsm32}
\eea
Now, to allow for no large unnatural cancellations in Eq.~\ref{eq:mzsm32}, we require 
$\mu^2\sim m_Z^2$
(same as $\Delta_{\rm EW}$) and also $a m_{3/2}^2\sim m_Z^2$. This latter condition can be fulfilled if
$m_{3/2}\sim m_Z$ (which now seems highly unlikely in light of LHC8 sparticle search limits and the value of $m_h$) 
{\it or} if $m_{3/2}$ is large but $a$ is small.
Since the $\mu$ term hardly evolves between $m_{\rm GUT}$ and $m_{\rm weak}$, we may equate
$-2m_{H_u}^2({\rm weak})\simeq am_{3/2}^2$. Since $am_{3/2}^2\sim m_Z^2$, then also $-m_{H_u}^2({\rm weak})\sim m_Z^2$: 
{\it i.e.} $m_{H_u}^2$ can start off large with magnitude of order $m_{3/2}$ at $m_{\rm GUT}$, but can be 
driven radiatively to small values $\sim -m_Z^2$ at $m_{\rm weak}$. 
This is the case of radiatively-driven naturalness.


\section{Naturalness in QCD: the need for axions}
\label{sec:qcd}

If we insist on naturalness in the electroweak sector, then it is only fair to 
insist as well on naturalness in the QCD sector.
In the early days of QCD, it was a mystery why the two-light-quark chiral symmetry U(2)$_L\times$U(2)$_R$
gave rise to three and not four light pions~\cite{U1}. 
The mystery was resolved by 't Hooft's discovery of the QCD theta vacuum which didn't respect 
the U(1)$_A$ symmetry~\cite{tHooft}. 
As a consequence of the theta vacuum, one expects the presence of a term 
\be
{\cal L}\ni \frac{\bar{\theta}}{32\pi^2}F_{A\mu\nu}\tilde{F}_A^{\mu\nu}
\ee 
in the QCD Lagrangian (where $\bar{\theta}=\theta+\arg(\det({\cal M}))$ and ${\cal M}$ 
is the quark mass matrix). Measurements of the neutron EDM constrain $\bar{\theta}\lesssim 10^{-10}$ 
leading to an enormous fine-tuning in $\bar{\theta}$: the so-called strong CP problem.

The strong CP problem is elegantly solved by Peccei, Quinn, Weinberg and Wilczek (PQWW)~\cite{pqww}
via the introduction of PQ symmetry and the concomitant (invisible~\cite{ksvz,dfsz}) axion: 
the offending term can dynamically settle to zero.
The axion is a valid dark matter candidate in its own right~\cite{axdm}.

Introducing the axion in a SUSY context solves the strong CP problem and 
renders naturalness to QCD. 
As a bonus, in the context of the SUSY DFSZ axion model~\cite{dfsz} 
where the Higgs superfields carry PQ charge,
one gains an elegant solution to the SUSY $\mu$ problem. 
The most parsimonius implementation of the strong CP solution
involves introducing a single MSSM singlet superfield $S$ carrying PQ charge $Q_{PQ}=-1$ while the
Higgs fields both carry $Q_{PQ}=+1$. The usual $\mu$ term is forbidden, but we have a 
superpotential~\cite{KN,susydfsz}
\be
W_{\rm DFSZ}\ni \lambda\frac{S^2}{M_P}H_uH_d .
\ee
If PQ symmetry is broken and $S$ receives a VEV $\langle S\rangle\sim f_a$, then a weak scale
$\mu$ term
\be
\mu\sim \lambda f_a^2/M_P
\ee
is induced which gives $\mu\sim m_Z$ for $f_a\sim 10^{10}$ GeV. Although Kim-Nilles sought to relate
the PQ breaking scale $f_a$ to the hidden sector mass scale $m_{\rm hidden}$~\cite{KN}, we see 
now that the Little Hierarchy 
\be
\mu\sim m_Z\ll m_{3/2}\sim {\rm multi-TeV} 
\ee
could emerge due to a mis-match between PQ breaking scale and hidden sector 
mass scale $f_a\ll m_{\rm hidden}$. For the remainder of this paper, we will assume
the SUSY DFSZ axion model holds due to its role in solving the SUSY $\mu$ problem.

\begin{figure}[tbp]
\begin{center}
\includegraphics[height=0.33\textheight]{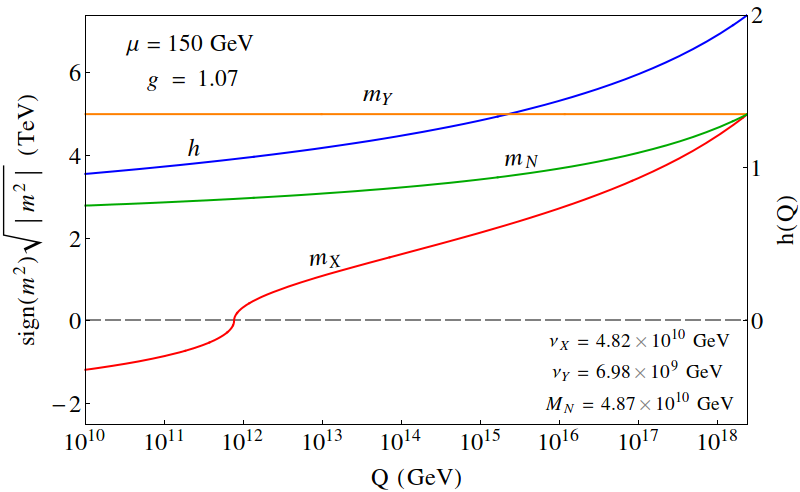}
\caption{Plot of the running values of PQ soft terms, set equal to $m_{3/2}=5$ TeV and $h=2$ at Planck scale versus $Q$.
Here, $g$ and $h$ are couplings from the MSY model Lagrangian~\cite{msy,radpq}.
\label{fig:run_m32}}
\end{center}
\end{figure}

An elegant model which exhibits this behavior was proposed by 
Murayama, Suzuki and Yanagida (MSY)~\cite{msy}. In the MSY model, PQ symmetry is broken
radiatively by driving one of the PQ scalars $X$ to negative mass-squared values in 
much the same way that electroweak symmetry is broken by radiative corrections 
driving $m_{H_u}^2$ negative.
Starting with multi-TeV scalar masses, the radiatively-broken PQ symmetry induces 
a SUSY $\mu $ term $\sim 100$ GeV~\cite{radpq} while at  the same time generating 
intermediate scale Majorana masses for right-hand neutrinos: see Fig.~\ref{fig:run_m32}.
Although we get different solutions for the PQ scale by setting $m_{3/2}$ to different masses at
Planck scale, the PQ scalar $X$ is driven to negative mass-squared values without 
$m_{3/2}$ dependence at the same $Q$ value. However, the coupling $h$ shifts the position of the $Q$ value where 
$m_X^{2}$ becomes negative; increasing $h$ shifts the point to higher energy scales.
In models such as MSY, the Little Hierarchy $\mu\ll m_{3/2}$ is no problem at all 
but is instead just a reflection of the mis-match between PQ and hidden sector mass scales.

\section{Relic abundance of axions and WIMPs with implications for axion detection}
\label{sec:axions}

\begin{figure}[tbp]
\begin{center}
\includegraphics[height=0.4\textheight]{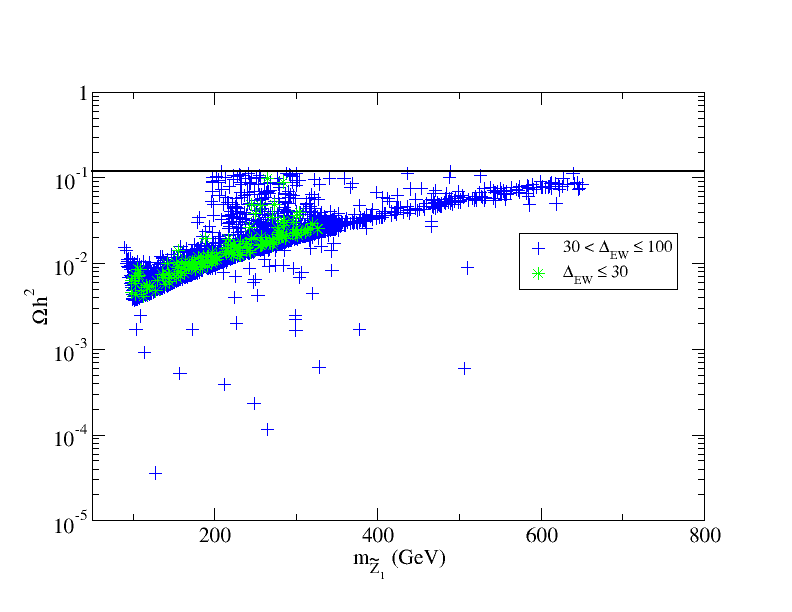}
\caption{Plot of standard thermal neutralino abundance 
$\Omega_{\tz_1}^{\rm std}h^2$ versus $m_{\tz_1}$ 
from a scan over NUHM2 parameter space with $\Delta_{\rm EW}<30$ (green stars) 
and $\Delta_{\rm EW}<100$ (blue crosses). 
We also show the central value of $\Omega_{\rm CDM}h^2$ from WMAP9.
\label{fig:oh2}}
\end{center}
\end{figure}

It is straightforward to calculate the thermal abundance of WIMPs in natural SUSY.
To a good approximation, it is given by 
\be
\Omega_{\tz_1}h^2= \frac{s_0}{\rho_c/h^2}\left(\frac{90}{\pi^2 g_*}\right)^{1/2}
\frac{x_{f}}{4M_{P}}
\frac{1}{\langle\sigma v\rangle}
\ee
where $s_0$ is the current entropy density of the universe, $\rho_c$ is the critical
density, $h$ is the scaled Hubble constant, $x_f=m_{\tz_1}/T_f$ is the scaled
WIMP freeze-out inverse temperature $\sim25$ and
$\langle\sigma v\rangle$ is the thermally averaged WIMP annihilation cross section
times relative velocity.
For a higgsino-like LSP as occurs in RNS, $\langle\sigma v\rangle$ is large
due to higgsino annihilation into vector boson pairs $WW$ and $ZZ$.
The simple ``WIMP miracle'' picture seems not to apply to higgsino dark matter
where we show in Fig.~\ref{fig:oh2} $\Omega_{\tz_1}h^2$ (from IsaReD~\cite{isared}) 
vs. $m_{\tz_1}$ 
from a scan over NUHM2 parameter space. 
Here, we keep only solutions with constraints: 1. $\mu >100$ GeV
in accord with LEP2 searches for chargino pair production, 2. $123<m_h<128$ GeV
in accord with the CERN Higgs discovery, and allowing for some theoretical error in the 
RG-improved one loop effective potential computation of $m_h$ in Isajet~\cite{isamh}
and 3. $\Delta_{\rm EW}<30$ (100) as denoted by green stars (blue crosses).
The plot shows that for low $m_{\tz_1}\sim 100$ GeV (as preferred by naturalness) 
the predicted thermal abundance of WIMPs is typically a factor 10-30
below the measured value of cold dark matter (CDM) $\Omega_{\rm CDM}h^2\simeq 0.12$. If we require
$\Delta_{\rm EW}<30$, then $m_{\tz_1}$ reaches $\sim 300$ GeV maximally with
$\Omega_{\tz_1}h^2$ as high as $0.02$. At some cost to naturalness, 
$\Omega_{\tz_1}h^2$ approaches the measured value for $m_{\tz_1}\sim 600$ GeV, 
where the $\tz_1$ is already frequently a mixed bino-higgsino particle.

Naively, one might expect natural SUSY to be ruled out as being incapable of 
generating a sufficiently large relic density of WIMPs. However, naturalness
in both EW and QCD sectors implies the presence of two dark matter particles:
the WIMP and the axion. Axions are expected to be produced dominantly via 
the non-thermal Bosonic Coherent Motion (BCM)~\cite{axdm,kimreview} yielding
\be
\Omega_a^{\rm std}h^2\simeq 0.23 f(\theta_i)\theta_i^2
\left(\frac{f_a/N_{\rm DW}}{10^{12}\ {\rm GeV}}\right)^{7/6}
\ee
where $\theta_i$ is the initial axion mis-alignment angle, $f_a$ is the axion decay
constant and $N_{\rm DW}$ is the domain-wall number. Also, $f(\theta_i)$ accounts for
anharmonicity effects. By proper choice of $f_a$ and $\theta_i$, BCM-produced axions
can always account for the measured CDM abundance.\footnote{Here, we impicitly assume that
PQ symmetry is broken before the end of inflation so that topological defects and archioles do not contribute
to the ultimate axion relic density\cite{sikivie,khlopov}.}

However, as mentioned previously, the axion superfield also contains a spin-1/2
axino $\ta$ and a spin-0 saxion $s$. In SUGRA, one expects $m_s\sim m_{3/2}$ while
the axino mass is more model-dependent but generally one expects also $m_{\ta}\sim m_{3/2}$~\cite{axino_mass}.
The DFSZ axinos can be produced thermally in the early universe  at a rate $\propto f_a^{-2}$
and largely independent of the re-heat temperature $T_R$~\cite{Bae:2011jb} 
(in the SUSY KSVZ model, then axino thermal production is proportional to $T_R$). Once axinos are produced, they undergo (late) decays
to sparticle plus particle thereby injecting additional WIMPs into the thermal plasma.
If enough WIMPs are produced at the axino decay temeprature, they undergo a process of
{\it re-annihilation} which still yields an enhanced WIMP abundance~\cite{az1}, 
but not as much as one-to-one with the population of thermally produced axinos. Of equal 
importance to WIMP production from axino decays is the axino decay temperature: if axinos 
decay before WIMP freeze-out, then the injected WIMPs thermalize and one regains the usual thermal WIMP abundance.
If axinos decay after WIMP freeze-out, then they always augment the WIMP abundance.

\begin{figure}
\begin{center}
\includegraphics[height=6.9cm]{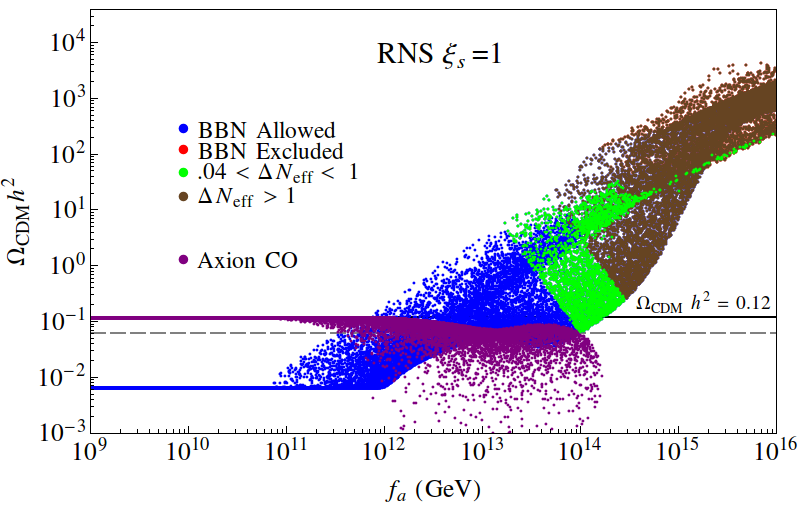}
\includegraphics[height=6.9cm]{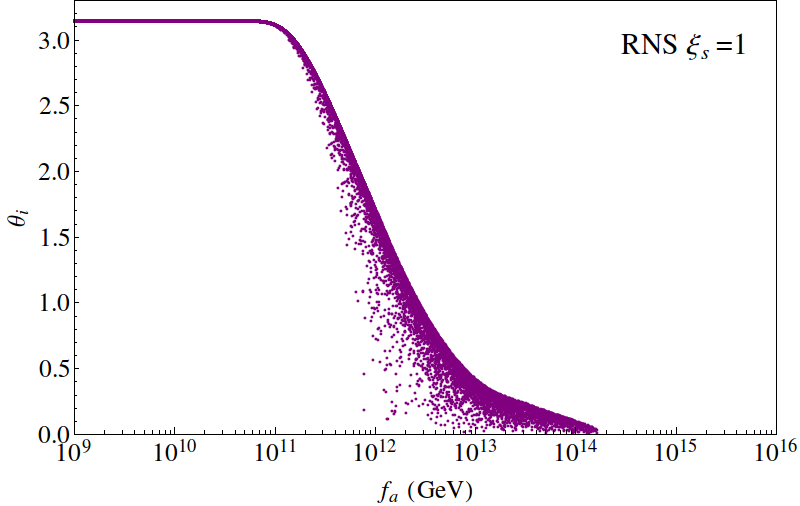}
\caption{In {\it a}) we plot the neutralino relic density from a scan over
SUSY DFSZ parameter space for the RNS benchmark case with $\xi=1$.
The grey dashed line shows the points where DM consists of 50\% axions and
50\% neutralinos.
The red BBN-forbidden points occur at $f_a\gtrsim 10^{14}$ GeV and are covered over by the brown $\Delta N_{\rm eff}>1$
coloration.
In {\it b}), we plot the misalignment angle $\theta_i$ needed to 
saturate the dark matter relic density $\Omega_{\tz_1 a}h^2=0.12$.
\label{fig:rns1}}
\end{center}
\end{figure}

Saxions can also be produced thermally at rates comparable to axino thermal production.
In addition, saxions can be produced via BCM which is especially important at large $f_a$. 
Since saxions are $R$-parity even, they can decay to pairs of SM particles, thereby injecting
extra entropy into the plasma, or they can decay to pairs of SUSY particles, thus
also augmenting the WIMP abundance (depending again on the saxion decay temperature).
Depending on a combination of PQ charge assignments and VEVs parametrized by $\xi_s$, 
the saxions may also decay to $\ta\ta$ (if kinematically allowed) thus adding to the WIMP
abundance, or they may decay to $aa$ thus injecting additional {\it dark radiation}
into the thermal plasma. Strong limits on dark radiation-- parametrized by the
effective number of additional neutrinos present in the universe $\Delta N_{\rm eff}$--
have been obtained, with a combination of Planck and other data sets finding
$N_{\rm eff}=3.15\pm0.23$~\cite{Planck:2015xua} (whereas the SM predicts $N_{\rm eff}=3.046$). 
Thus, too much dark radiation from saxion decay can lead to conflict with measured cosmological parameters.
In our numerical study, we consider a conservative constraint $\Delta N_{\rm eff}<1$ (see Fig.~\ref{fig:rns1}) 
at over $3 \sigma$ with the joint $Planck$ TT+lowP+BAO result~\cite{Planck:2015xua}.
In addition, if saxions or axinos of sufficient initial abundance decay after the onset of
BBN, then they can destroy the successful predictions of light element abundances 
via BBN, and again the model can be excluded.
\begin{figure}
\begin{center}
\includegraphics[height=9cm]{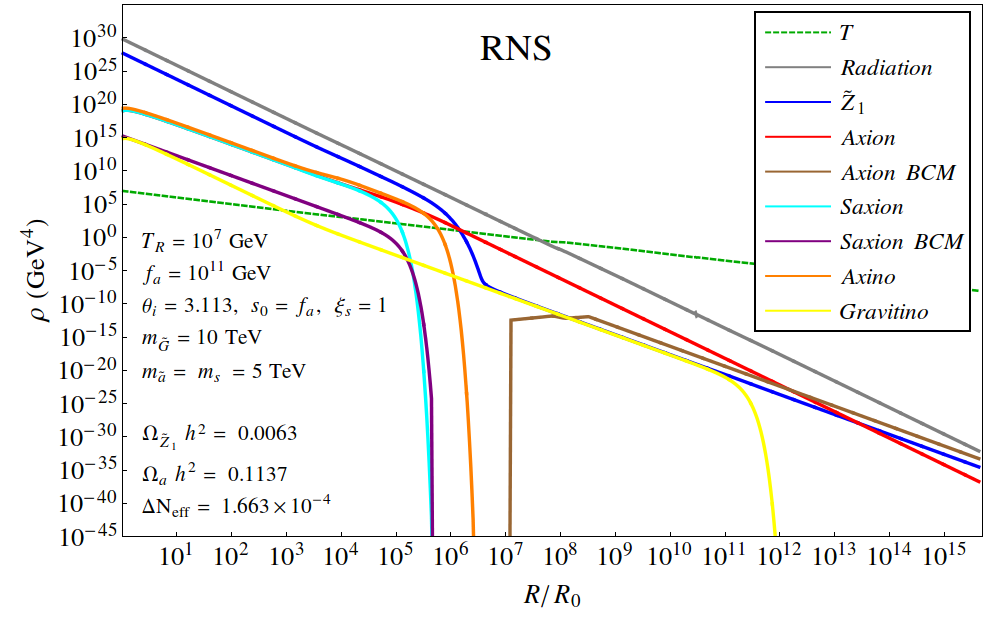}
\caption{Evolution of various energy densities vs. scale factor $R/R_0$
for the RNS benchmark case with $\xi_s=1$ and other parameters as indicated
in the figure.
\label{fig:rho1}}
\end{center}
\end{figure}

The calculation of the mixed axion-neutralino relic abundance can be calculated
via semi-analytic techniques~\cite{dfsz1} or more reliably~\cite{dfsz2}
via the simultaneous solution of eight coupled Boltzmann equations describing
the energy densities of 1. radiation, 2. thermally- and decay-produced WIMPs, 3.
BCM-produced axion, 4. BCM produced saxions (followed by saxion decay), thermal production and decay of 5. axino, 6. saxions and 7. thermal and decay-induced production of axions
and 8. thermal production and decay of gravitinos.

We scan over the following PQ parameters:
\bea
 10^9 \mbox{ GeV} < & f_a & < 10^{16} \mbox{ GeV}, \nonumber \\
 0.4 \mbox{ TeV} < & m_{\ta} & < 20 \mbox{ TeV},\\
 0.4 \mbox{ TeV} < & m_s & < 20 \mbox{ TeV}. \nonumber
\eea
The result of these calculations were shown in Ref.~\cite{dfsz2} and in Fig.~\ref{fig:rho1}
where the energy densities are tracked as a function of scale factor $R$ from the end of
inflation with $T=T_R$ to the era of entropy conservation.
\begin{figure}
\begin{center}
\includegraphics[height=6.9cm]{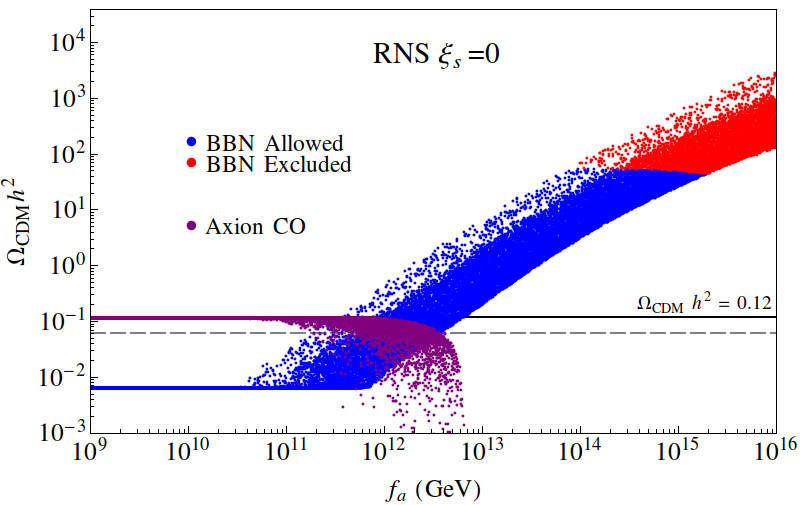}
\includegraphics[height=6.9cm]{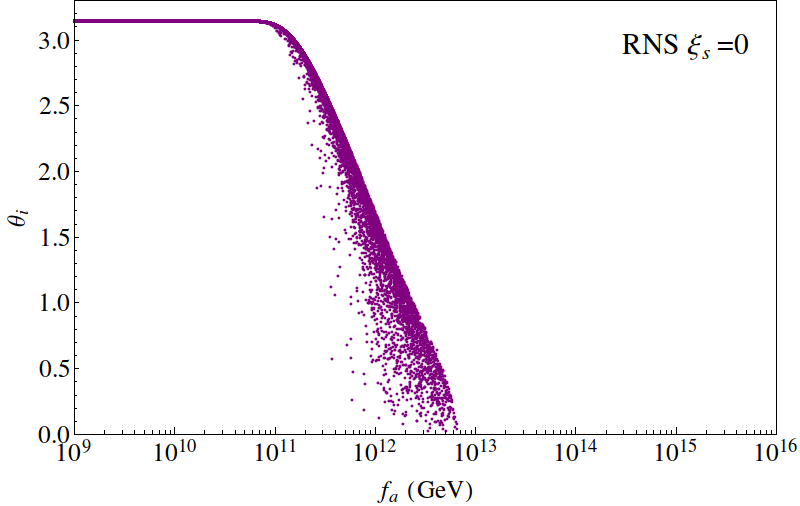}
\caption{In {\it a}) we plot the neutralino relic density from a scan over
SUSY DFSZ parameter space for the RNS benchmark case with $\xi_s=0$.
The grey dashed line shows the points where DM consists of 50\% axions and
50\% neutralinos.
In {\it b}), we plot the misalignment angle $\theta_i$ needed to 
saturate the dark matter relic density $\Omega_{\tz_1 a}h^2=0.12$.
\label{fig:rns0}}
\end{center}
\end{figure}

In Fig.~\ref{fig:rns0}, we show the calculated relic abundance of both WIMPs (blue and red points)
and axions (purple points) as a function of $f_a$ for a RNS benchmark SUSY model 
with $m_0=5000$ GeV, $m_{1/2}=700$ GeV, $A_0=-8300$ GeV, $\tan\beta =10$, $\mu =110$ GeV
and $m_A=1000$ GeV. We first take $\xi_s=0$ so saxion decays to $aa$ and $\ta\ta$ are turned off.

At very low $f_a$, axinos are thermally-produced at a large rate but also decay well before
neutralino freeze-out so that the WIMP abundance is still given by its
expected thermally-produced value. As $f_a$ increases, ultimately axinos begin decaying
after freeze-out thus augmenting the WIMP abundance. For $f_a>10^{13}$ GeV, too many WIMPs
are produced and the model parameters are excluded. For very large $f_a\sim 10^{15}$ GeV, 
all points are doubly excluded by producing too much dark matter and violating limits 
from BBN~\cite{jedamzik}. We also show the axion abundance. 
At very low $f_a$, the CDM is axion-dominated~\cite{dfsz1} although this requires very high values 
of $\theta_i\sim \pi$ (see frame~\ref{fig:rns0}{\it b}). which might be considered fine-tuned. 
For $f_a\sim 10^{12}$ GeV, axions can still
dominate the CDM abundance but with $\theta_i\sim 1$. For these values of $f_a$, the CDM could also
easily be WIMP dominated as well.

In Fig.~\ref{fig:rns1}, we show the neutralino and axion relic abundance for the RNS
benchmark with $\xi_s=1$ (saxion decays to axions and axino pairs are turned on).
In this case, the additional decay modes allow the saxion to be shorter lived 
for a given value of $m_s$ and $f_a$ compared to the $\xi_s=0$ case. As a consequence, there is a greater range of $f_a$ where CDM can be axion-dominated. Ultimately, axinos and saxions decay
after freeze-out and the WIMP abundance is enhanced at higher $f_a\sim 10^{11}-10^{14}$ GeV.
For $f_a\gtrsim 10^{14}$ GeV, WIMPs are overproduced. Points at very high $f_a$ for $\xi_s=1$
can be triply excluded by producing too many WIMPs and by violating both 
dark radiation and BBN constraints.
\begin{figure}
\begin{center}
\includegraphics[height=4.6cm]{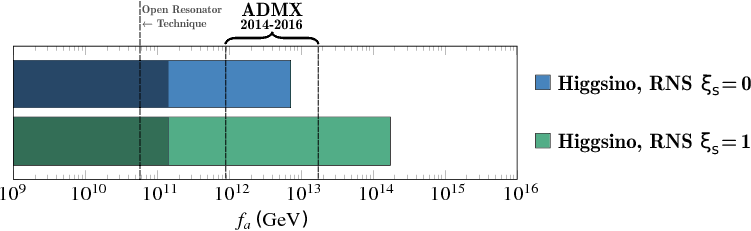}
\caption{Range of $f_a$ which is allowed in each PQMSSM scenario for the 
RNS benchmark models. 
Shaded regions indicate the range of $f_a$ where $\theta_i>3$.
\label{fig:bar}}
\end{center}
\end{figure}

We summarize the results of this section in Fig.~\ref{fig:bar}. We display the range of 
$f_a$ where valid solutions for the relic abundance of mixed axion-higgsino CDM can be found
for the RNS benchmark model. The upper bar shows the range of $f_a$ for $\xi_s=0$ while the
lower bar shows the range for $\xi_s=1$. The darker shaded parts of the bars denote 
$\theta_i$ values $>3$ which might be considered less plausible or fine-tuned.
We also show by the bracket the range of $f_a$, assuming the bulk of DM is axion, 
which is expected to be probed by the ADMX experiment within the next several years~\cite{admxtalk}. 
This region probes the most natural region
where $\theta_i\sim 1$. We also show a further region of lower $f_a$ which might be explored by
a new open resonator technology~\cite{Rybka:2014cya}. About a decade of natural $f_a\sim 10^{14}$ GeV 
seems able to elude ADMX searches for the $\xi_s=1$ case.

\subsection{Results for variable $\mu$}

We may convert the RNS benchmark point into a model line by allowing for variable $\mu$.
In this case, we have variable higgsino mass with the lower bound given by 
the LEP2 limit $\mu\sim m_{\tw_1}>103.5$ GeV while the upper bound is determined
by how much fine-tuning one is willing to tolerate with 
\be 
\mu^2 <\Delta_{\rm EW}^{\rm max} m_Z^2/2 .
\ee
\begin{figure}
\begin{center}
\includegraphics[height=7cm]{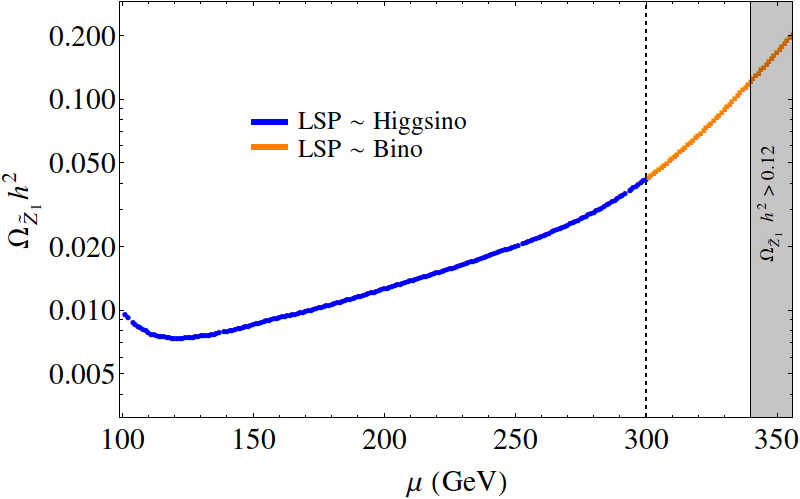}
\caption{Thermally-generated neutralino abundance vs. $\mu$
for the RNS benchmark model-line. Vertical dashed line shows the point where the $\tz_1$ 
becomes more bino-like than higgsino-like (or vice versa).
\label{fig:omg_mu}}
\end{center}
\end{figure}

In Fig.~\ref{fig:omg_mu}, we show the thermally-produced relic density
of neutralinos along the variable $\mu$ RNS model line. Here, $\Omega_{\tz_1}h^2\sim 0.007$ 
for low $\mu$ but increases as $\mu$ increases since the $\tz_1$ becomes
increasingly bino-like. At $\mu\sim 300$ GeV, the $\tz_1$ becomes more
bino-like than higgsino-like, and at $\mu\sim 340$ GeV, naively too much
neutralino dark matter is produced. As we have seen, it is easy to increase the neutralino 
abundance from its thermal expectation by allowing for axino and saxion production
with decay taking place after neutralino freeze-out. It is much harder to {\it reduce}
the neutralino abundance from its thermal value: the three most common ways include
1. entropy dilution from saxion decay to SM particles only at very high $f_a\sim 10^{15}$ GeV, 
2. allowing for $R$-parity violation (in which case one must somehow stabilize the proton) or
3. allowing for a lighter LSP than the neutralino ({\it e.g.} a light axino or gravitino
into which the neutralino may decay).
\begin{figure}
\begin{center}
\includegraphics[height=7cm]{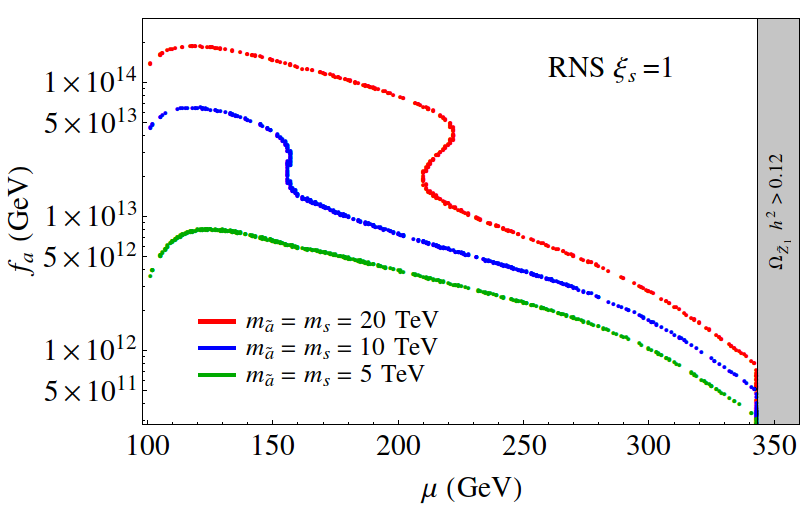}
\caption{Contours of allowed $f_a$ region as $\mu$ is varies along the
RNS benchmark model-line. 
\label{fig:fa_bound}}
\end{center}
\end{figure}

In Fig.~\ref{fig:fa_bound}, we plot the contours of allowed regions (allowed below the contours) 
in the  $f_a$ vs. $\mu$ plane by varying $\mu$ along the RNS model line for the $\xi_s=1$ case. 
We show the boundaries for three different assumptions on the axino/saxion masses: $m_{\ta,s}=5$, 10 and 20 TeV.
The lower bound is always $f_a\gtrsim 10^9$ GeV from supernovae/red giant astrophysical
cooling limits~\cite{redgiant} (although the lower range requires some tuning 
on $\theta_i\sim \pi$). For our canonical case where we expect
$m_0\simeq m_{\ta,s}\sim m_{3/2}=5$ TeV, $f_a$ can range up to $10^{13}$ GeV
beyond which too much neutralino mass density is produced. As $\mu$ increases, the upper
bound tends to decrease because the neutralino thermal abundance is increasing
and there is less ``room'' for additional neutralino production from axino/saxion decay.
As $m_{\ta,s}$ increase, the upper bound on $f_a$ increases. This is because
as $m_{\ta,s}$ become more massive, their widths increase and their lifetimes decrease: 
for a given $f_a$ value, they are more likely to decay at earlier times and so 
re-annihilation from decay-produced neutralinos occurs at higher axino/saxion
decay temperature (and the re-annihilation yield is inversely proportional to
decay temperature~\cite{az1}). For the $\xi_s=0$ case, we have more constrained upper 
$f_a$ boundaries since saxion decays into axions and axinos are turned off and 
hence the saxion is longer lived.

\section{Direct and indirect detection of WIMPs}
\label{sec:wimps}

In this Section, we update our previous projections~\cite{bbm} for direct and indirect detection of
higgsino-like WIMPs from radiatively-driven natural SUSY.
Our current results contain several improvements:
\begin{enumerate}
\item Our previous scan over NUHM2 parameter space was restricted to a range of $m_A:0.15-1.5$ TeV.
However, low $\Delta_{\rm EW}$ solutions can be found for much higher $m_A$ values~\cite{rnshiggs} and
so here we expand the $m_A$ range to as far as 20 TeV so that the bounds on our scanned
parameter space is dictated by the value of $\Delta_{\rm EW}$ rather than an arbitrary parameter cutoff.
\item We have updated the nucleon mass fraction parameters which enter the quark and gluon matrix elements
in IsaReS~\cite{isares} to values given in Table 1 of Ref.~\cite{hisano}. 
These mainly lessen the contribution from strange quarks
from older estimates of the spin-dependent scattering cross section. In our case, the
computed values of $\sigma^{\rm SI}(\tz_1 p)$ decrease by typically a factor of two.
\item We have increased our sampling statistics in NUHM2 parameter space.
\end{enumerate}
\begin{figure}[tbp]
\begin{center}
\includegraphics[height=0.4\textheight]{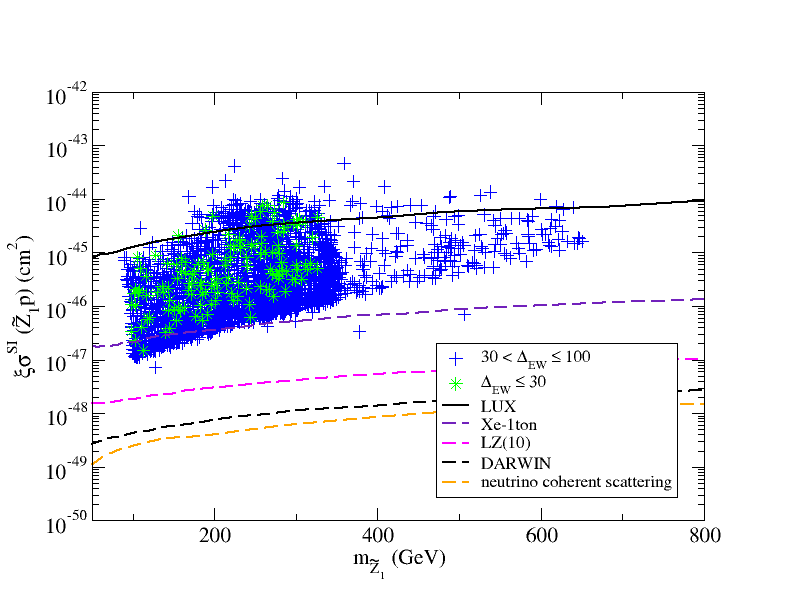}
\caption{Plot of rescaled higgsino-like WIMP spin-independent 
direct detection rate $\xi \sigma^{\rm SI}(\tz_1 p)$ 
versus $m_{\tz_1}$ from a scan over NUHM2 parameter space with $\Delta_{\rm EW}<30$ (green)
and $30<\Delta_{\rm EW}<100$ (blue). 
We also show the current reach from the LUX experiment
and projected reaches of Xe-1-ton, LZ(10) and Darwin.
\label{fig:SI}}
\end{center}
\end{figure}

In Fig.~\ref{fig:SI}, we show the spin-independent neutralino-proton scattering rate in $\rm cm^2$
as calculated using the updated IsaReS~\cite{isares}. 
The result is rescaled by a factor $\xi=\Omega_{\tz_1}^{\rm std}h^2/0.12$
to account for the fact that the local relic abundance might be less than the usually assumed value
$\rho_{\rm local}\simeq 0.3$ GeV/cm$^3$, as suggested long ago by Bottino {\it et al.}~\cite{bottino} 
(the remainder would be composed of axions). 
Green stars denote points with $\Delta_{\rm EW}<30$ while blue crosses denote points with $30<\Delta_{\rm EW}<100$.

The higgsino-like WIMP in our case scatters from quarks and gluons mainly via $h$ exchange. 
The $\tz_1 -\tz_1 -h$ coupling involves a product of both higgsino and gaugino components. In the case of RNS models, 
the $\tz_1$ is mainly higgsino-like, but since $m_{1/2}$ is bounded from above by naturalness, the $\tz_1$
contains enough gaugino component that the coupling is never small: in the notation of Ref.~\cite{wss}
\be
{\cal L}\ni -X_{11}^h \overline{\tz}_1 \tz_1 h
\ee
where
\be
X_{11}^h =-{1\over 2}\left(v_2^{(1)}\sin\alpha -v_1^{(1)}\cos\alpha \right) 
\left(gv_3^{(1)}-g'v_4^{(1)}\right) ,
\ee
and where $v_1^{(1)}$ and $v_2^{(1)}$ are the higgsino components and $v_3^{(1)}$ and $v_4^{(1)}$ are
the bino and wino components of the lightest neutralino, $\alpha$ is the Higgs mixing angle and $g$ and 
$g^\prime$ are SU(2)$_L$ and U(1)$_Y$ gauge couplings.
Thus, for SUSY models with low $\Delta_{\rm EW}\lesssim 30-100$, the SI direct detection cross section is also 
bounded from below, even including the rescaling factor $\xi$.

From Fig.~\ref{fig:SI}, we see that the current reach from the LUX experiment (solid contour)
has begun sampling the upper limits of predicted $\xi \sigma^{\rm SI}(\tz_1 p)$ values.
The projected reach of Xe-1-ton, a ton scale
liquid Xenon detector, is also shown. 
It is seen to cover nearly all the predicted parameter space points.
We also show the projected reach of LZ(10), an upgrade to LUX.
The projected LZ reach is seen to cover the entire set of points generated.
Thus, the projected ton scale noble liquid detectors (or other comparable WIMP detectors) 
can make a {\it complete} exploration of the RNS parameter space. 
Since deployment of these ton-scale detectors is ongoing,
it seems that direct WIMP search experiments may either verify or exclude RNS models in the near future.
These searches should either verify or rule out a very essential aspect of natural SUSY models.
\begin{figure}[tbp]
\begin{center}
\includegraphics[height=0.4\textheight]{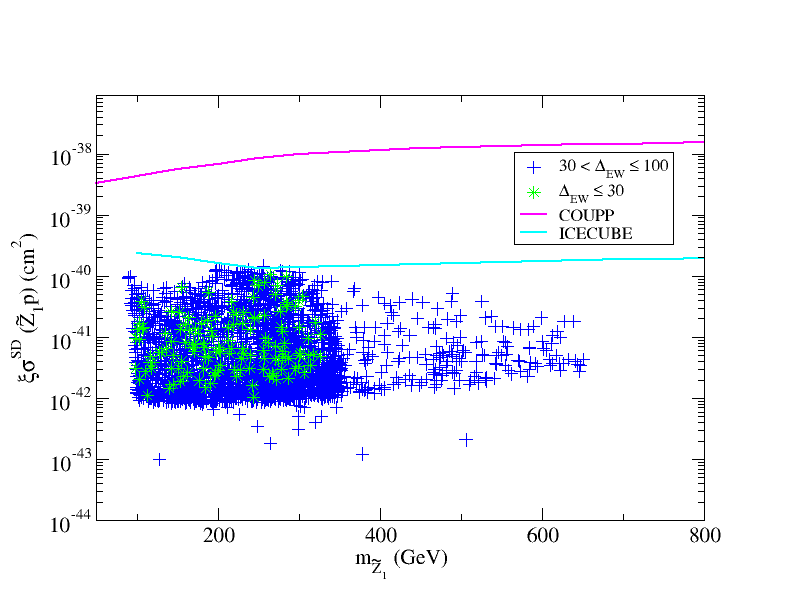}
\caption{Plot of rescaled spin-dependent higgsino-like WIMP detection rate $\xi \sigma^{\rm SD}(\tz_1 p)$ 
versus $m_{\tz_1}$ from a scan over NUHM2 parameter space with $\Delta_{\rm EW}<30$ (green stars)
and $30<\Delta_{\rm EW}<100$ (blue crosses). 
We also show current reach from the COUPP and IceCube detectors.
\label{fig:SD}}
\end{center}
\end{figure}

In Fig.~\ref{fig:SD}, we show the rescaled {\it spin-dependent} neutralino-proton scattering
cross section $\xi\sigma^{\rm SD}(\tz_1 p)$. Here we show recent limits from the COUPP~\cite{coupp} 
detector. Current limits are still about an order of magnitude away from reaching the
predicted rates from RNS models.
We also show limits from the IceCube experiment. IceCube searches for high energy neutrinos which 
could be produced from WIMP annihilations in the solar core. The IceCube expected rates depend on the Sun's ability
to capture WIMPs which in turn depends on a product of spin-dependent neutralino-proton scattering cross section
times the local WIMP abundance.\footnote{In a previous work~\cite{bbm}, it was mistakenly suggested that the
IceCube detection rate was independent of local abundance due to equilibration between solar capture rate and WIMP annihilation rate.}
The IceCube limits have barely entered the RNS parameter space and excluded just the largest values
of $\xi \sigma^{\rm SD}(\tz_1 p)$. 

%
%

%
\begin{figure}[tbp]
\begin{center}
\includegraphics[height=0.4\textheight]{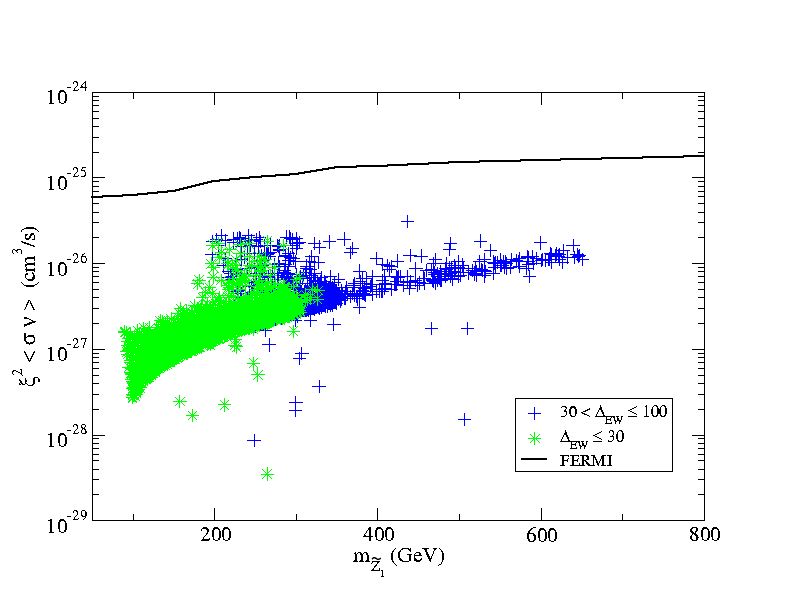}
\caption{Plot of rescaled $\xi^2 \langle\sigma v\rangle |_{v\to 0}$ 
versus $m_{\tz_1}$ from a scan over NUHM2 parameter space with $\Delta_{\rm EW}<30$ (green stars)
and $30<\Delta_{\rm EW}<100$ (blue crosses). 
We also show current reach from Fermi LAT, 
Ref.~\cite{fermi}.
\label{fig:sigv}}
\end{center}
\end{figure}

In Fig.~\ref{fig:sigv}, we show the rescaled thermally-averaged neutralino annihilation
cross section times relative velocity in the limit as $v\to 0$: $\xi^2\langle\sigma v\rangle|_{v\to 0}$.
This quantity enters into the rate expected from WIMP halo annihilations into
$\gamma$, $e^+$, $\bar{p}$ or $\bar{d}$. 
The rescaling appears as $\xi^2$ since limits depend on the square of the local WIMP abundance~\cite{bottino_id}.
Anomalies in the positron and $\gamma$ spectra
have been reported, although the former may be attributed to pulsars~\cite{pulsars}, 
while the latter 130 GeV gamma line may be instrumental. 
On the plot, we show the limit derived from 
the Fermi LAT gamma ray observatory~\cite{fermi} for WIMP annihilations into $WW$. 
These limits have not yet reached the RNS parameter space due in part to suppression from 
the squared rescaling factor.


\section{Conclusions}
\label{sec:conclude}

We have found in this paper, and in previous works, that if one insists
on naturalness-- in both the electroweak and the QCD sectors-- then the
simple picture of SUSY WIMP dark matter changes radically. 
Naturalness in the electroweak sector implies a low value of the 
superpotential $\mu$ parameter: the closer to $m_Z$ the better. In models with gaugino mass 
unification, as favored in simple GUTs, this implies the LSP is a higgsino-like
neutralino with a predicted thermal abundance a factor of 10-15 below the measured
dark matter density. This seeming disaster is in fact an attribute if one also insists
on naturalness in the QCD sector, {\it i.e} solving the strong CP problem. 
In this case, the most compelling solution invokes a PQ symmetry with its concommitant axion.
In this situation, the axion makes up the remaining abundance, and in fact over most of
parameter space the axion is the dominant CDM component while WIMPs are subdominant. 

Invoking the axion in a SUSY context brings along both the axino and the saxion. 
The dark matter abundance calculation becomes more intertwined since axions can be produced
via BCM, via thermal production and via saxion decay. WIMPs can be produced thermally but also
via axino, saxion and gravitino decays. If WIMPs are produced via decays 
at sufficient rates, then WIMP re-annihilation occurs. Additional entropy can be produced
at late times by the decays of heavy unstable states, thus diluting all relics which are present.
The ensuing abundance calculation is more complicated than the simple WIMP miracle picture, but
in many ways it is more elegant and compelling. Our abundance calculations here have
used the SUSY DFSZ axion model which provides an elegant solution to the SUSY $\mu$ problem.
We have outlined the range of $f_a$ values which are allowed in RNS, and shown the
regions which ADMX and other experiments hope to probe in the near future.

With regard to WIMP detection, higgsino-like LSPs which contain significant gaugino 
components (as is required in natural SUSY) generally have large rates for both
direct and indirect detection, at least compared to binos. However, the propitious 
detection rates are off-set by the fact that now the WIMPs might 
comprise only a small fraction of the local abundance instead of the entirety of CDM. 
To compensate, one must temper detection rates by the $\xi=\Omega_{\tz_1}h^2/0.12$ 
factor. For instance, direct detection via SI or SD scattering are both reduced by a factor
$\xi$. Nonetheless, ton-scale noble liquid WIMP detectors are projected to probe the
{\it entirety} of RNS parameter space: if a WIMP signal is not ultimately seen, then
the RNS picture will have to be seriously modified or abandoned.
Detection rates for indirect WIMP searches
via halo WIMP annihilation into gammas or antimatter are suppressed by a factor of $\xi^2$.
This suppression will make detection of WIMPs in these channels more difficult, except in the cases
where WIMPs still comprise the bulk of dark matter.


\acknowledgments{Acknowledgments}

We thank X. Tata, A. Mustafayev, A. Lessa, M. Padeffke-Kirkland, D. Mickelson, E. J. Chun 
and P. Huang for collaborations and discussions leading to the results presented here.
The computing for this project was performed at the 
OU Supercomputing Center for Education \& Research (OSCER) at the University of
Oklahoma (OU).
We thank D. Cline for soliciting this manuscript and for financial assistance.
This work is supported in part by the US Department of Energy Office of High Energy Physics.







\bibliographystyle{mdpi}
\makeatletter
\renewcommand\@biblabel[1]{#1. }
\makeatother



%


%

\end{document}